\documentclass[fleqn,10pt]{elsarticle}
\usepackage{lineno}
\usepackage{multirow}
\usepackage{subfigure}
\usepackage{graphicx}
\usepackage{graphics}
\usepackage{epstopdf}
\usepackage{extarrows}
\usepackage{xcolor}
\usepackage{geometry}
\usepackage{url}
\usepackage{float}

\usepackage{mathrsfs}
\usepackage{subfigure}
\usepackage{amssymb}
\usepackage{amsmath}
\usepackage{color}
\usepackage{ulem}
\usepackage{array}
\usepackage{longtable}
\usepackage{lscape}
\usepackage{multicol}
\numberwithin{equation}{section}
\allowdisplaybreaks

\modulolinenumbers[5]

\begin{document}

\renewcommand{\vec}[1]{\boldsymbol{#1}}
	
\begin{frontmatter}
	
\title{A computational scheme connecting gene regulatory network dynamics with heterogeneous stem cell regeneration}
	
\author[add1]{Yakun Li}
\author[add1]{Xiyin Liang}
\author[add1]{Jinzhi Lei\corref{correspondingauthor}}
	
\address[add1]{
School of Mathematical Sciences, 
Center for Applied Mathematics, 
Tiangong University, 
Tianjin, 300387, China
}

\cortext[correspondingauthor]{Corresponding author:  jzlei@tiangong.edu.cn (J. Lei)}
	
\begin{abstract}
Stem cell regeneration is a vital biological process in self-renewing tissues, governing development and tissue homeostasis. Gene regulatory network dynamics are pivotal in controlling stem cell regeneration and cell type transitions. However, integrating the quantitative dynamics of gene regulatory networks at the single-cell level with stem cell regeneration at the population level poses significant challenges. This study presents a computational framework connecting gene regulatory network dynamics with stem cell regeneration through a data-driven formulation of the inheritance function. The inheritance function captures epigenetic state transitions during cell division in heterogeneous stem cell populations. Our scheme allows the derivation of the inheritance function based on a hybrid model of cross-cell-cycle gene regulation network dynamics. The proposed scheme enables us to derive the inheritance function based on the hybrid model of cross-cell-cycle gene regulation network dynamics. By explicitly incorporating gene regulatory network structure, it replicates cross-cell-cycling gene regulation dynamics through individual-cell-based modeling. The numerical scheme holds the potential for extension to diverse gene regulatory networks, facilitating a deeper understanding of the connection between gene regulation dynamics and stem cell regeneration.
\end{abstract}


\end{frontmatter}
	
\section{Introduction}
\label{sec:level1}
	
Stem cell regeneration is a critical biological process in the development and maintenance of tissues, ensuring their self-renewal and homeostasis\cite{clevers2015adult}. The intricate interplay between gene expression regulation and cell population dynamics is central to maintaining dynamic equilibrium during stem cell regeneration. However, as these processes are typically described by distinct mathematical models, bridging the gap between their scales of dynamics is imperative. 

In the context of heterogeneous stem cell regenerations, a comprehensive mathematical framework has emerged, integrating cell division behavior with cell heterogeneity and random transition of epigenetic states\cite{lei2020general,Lei2020}. This framework introduces a variable $\vec{x}$, often a high-dimensional vector, to represent the epigenetic state of a cell. The evolution of the population of cells with epigenetic state $\vec{x}$, denoted as $Q(t, \vec{x})$, is governed by a differential-integral equation\cite{lei2020general,Lei2020}
\begin{equation}
	\label{4}
	\left\{
	\begin{aligned}
		\frac{\partial Q(t, \vec{x})}{\partial t} =&
		-Q(t,\vec{x})(\beta(c(t), \vec{x})+\kappa(\vec{x}))\\
		&+2\int _{\Omega} \beta (c(t-\tau(\vec{y})), \vec{y}) Q(t-\tau( \vec{y}), \vec{y})e^{-\mu( \vec{y})\tau( \vec{y})}p(\vec{x}, \vec{y}) d \vec{y},\\
		c(t)=&\int_{\Omega} Q(t, \vec{x})\zeta(\vec{x})d \vec{x}.\\
	\end{aligned}
	\right.
\end{equation}
Here, $\beta$, $\kappa$, and $\mu$ represent the rates of cell proliferation, differentiation/senescence, and apoptosis, respectively; $\tau$ indicates the duration of cell proliferation; $\zeta$ signifies the rate of cytokine secretion. The concentration of growth factors secreted by all cells is denoted by $c$, and $p(\vec{x}, \vec{y})$ quantifies the probability of transferring epigenetic states during cell division. The equation \eqref{4} was proposed by generalizing the classical G0 cell cycle model and incorporating the epigenetic state of individual cells represented by a continuous multidimensional variable\cite{lei2020general}. 

The epigenetic state $\vec{x}$ in \eqref{4} represents intrinsic cellular states that may dynamically change over time within a cell cycle or during cell division. Biologically, the epigenetic state of a cell can refer to molecular level changes that are independent of the DNA sequences, including DNA methylation, nucleosome histone modifications, and gene expression\cite{Probst2009,Schepeler:2014ir,SerraCardona:2018bz,Singer:2014eua,Takaoka:2014gg,Wu:2014gw}. The epigenetic state may influence the quantification of cellular behaviors, such as proliferation ($\beta$), differentiation ($\kappa$), apoptosis ($\mu$), and cell cycle ($\tau$), which collectively define the \textit{kinetotype} of a cell\cite{lei2020general}. 

Gene regulatory dynamics within a cell cycle are often described by a set of ordinary or stochastic differential equations
\begin{equation}
	\label{1}
	\frac{d \vec{X}}{dt} = \vec{F}(\vec{X}) + \vec{\eta}(t),
\end{equation}
where $\vec{X}$ represents the vector of gene expression. The function $\vec{F}$ describes the regulatory relationship defined by gene regulatory networks, and $\vec{\eta}(t)$ denotes the stochastic fluctuations in molecular concentration changes. When the epigenetic state $\vec{x}$ is associated with gene expression levels $\vec{X}$, e.g., $\vec{x} = \log (1 + \vec{X})$, equation \eqref{1} describes the dynamics of the epigenetic state inside a single cell.

In addition to gene regulatory dynamics within a cell cycle, epigenetic states undergo random changes during cell division, leading to cell plasticity in regeneration. This randomness is encapsulated by the inheritance function $p(\vec{x}, \vec{y})$ in \eqref{4}, which plays a crucial role in characterizing cell type changes. However, precisely formulating this function poses challenges due to the complexity of biochemical reactions during cell divisions. 

Biologically, $p(\vec{x}, \vec{y})$ represents the probability that a daughter cell with state $\vec{x}$ originated from a mother cell with state $\vec{y}$ after division. Thus, treating $p(\vec{x}, \vec{y})$ as a conditional probability density
\begin{equation}
\label{eq:29}
p(\vec{x}, \vec{y}) = P(\mbox{state\ of\ dauther\ cell} = \vec{x}\ \vert\ \mbox{state\ of\ mother\ cell} = \vec{y}),
\end{equation}   
we can focus on the epigenetic state before and after cell division while simplifying the analysis of the intermediate processes. The mathematical formulation of the inheritance function is often difficult to derive following the complex biological process. Numerical simulations based on epigenetic state inheritance laws\cite{huang2018dynamics,huang2018cell,sahoo2021mechanistic} offer phenomenological representations of the inheritance function. For instance, normalized nucleosome modifications follow a beta distribution\cite{huang2018dynamics}, while transcript levels follow a gamma distribution\cite{cai2006stochastic}. Nevertheless, how the structure of the gene regulation network may be involved in the inheritance function remains not known. 
		
This study aims to develop a computational scheme linking gene regulatory network dynamics and stem cell regeneration through a data-driven formulation of the inheritance function. We focused on a well-studied gene network governing cell fate decisions and cell type switches, devising a numerical scheme to obtain the conditional probability density $p(\vec{x}, \vec{y})$. The obtained density function, verified through stochastic simulation, provides insights into how gene regulatory network structure influences inheritance. Importantly, our numerical scheme offers a pathway for extending these findings to broader gene regulatory networks, facilitating deeper insights into the nexus between gene regulation dynamics and stem cell regeneration. 
	
\section{Model and Method}
\label{sec:level}
	
We referred to the hybrid model of stem cell differentiation developed to study the dynamics of cell-type transitions mediated by epigenetic modifications\cite{huang2023dynamics}. This model integrates individual-cell-based modeling of a multicellular system, dynamics of a gene regulation network (GRN) involving two genes, a G0 cell cycle model governing cell regeneration, and stochastic inheritance of epigenetic states during cell divisions. 
	
\subsection{Hetereogeneous G0 cell cycle model}
 	
\begin{figure}[htbp]
 	\centering
 	\includegraphics[width=10cm]{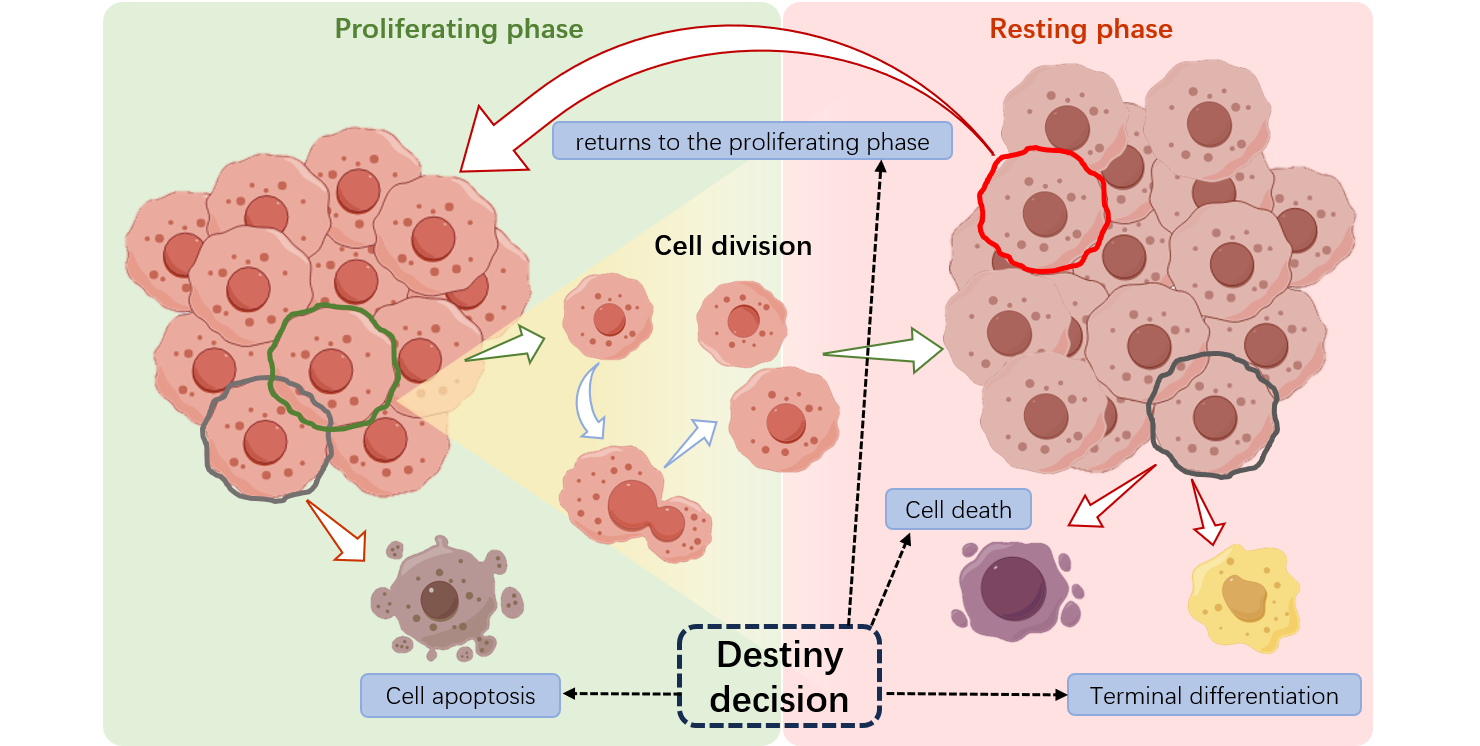}
 	\caption{\textbf{G0 cell cycle model of cell regeneration}. Schematic representation of the G0 model depicting stem cell regeneration. Cells in the resting phase either enter the proliferative phase at a rate $\beta$ or exit the resting pool at a rate $\kappa$ due to differentiation, senescence, or death. Proliferating cells undergo apoptosis at a rate $\mu$.}
 	\label{fig:2}
\end{figure}

In our model, we focused solely on cells capable of cycling, excluding those that had lost this capacity from the simulation pool. We utilized the G0 cell cycle model to describe the dynamics of stem cell regeneration\cite{Burns:1970tm,Mackey:1978vy}. According to this model, cells progress through a resting phase known as G0, during which they undergo growth and prepare to enter the proliferative phase upon receiving cell cycling checkpoint signals (Fig. \ref{fig:2}). Each cell in the resting phase has the potential to either enter the proliferating phase at a rate $\beta$ or exit the resting pool at a rate $\kappa$ due to processes such as terminal differentiation, senescence, or death. Cells in the proliferating phase are random eliminated at a rate $\mu$ due to apoptosis or undergo mitosis at a fixed time $\tau$ following entry into the proliferative compartment. Mitosis results in the generation of two daughter cells from each mother cell. The newborn cells then enter the resting phase, initiating the next cycle.

Biologically, the self-renewal capability of a cell is intricately linked to microenvironmental conditions, including growth factors, various cytokines, and intracellular signaling pathways\cite{Moustakasetal.2002,Yangetal.2010,Ornitz:2001vh}. Despite the complexity of these pathways, the phenomenological formation of Hill function dependence can be derived from simple assumptions regarding interactions between signaling molecules and receptors\cite{lei2020general,Bernard:2003ct}, expressed as:
\begin{equation}
\beta = \beta_0 \dfrac{\theta^n}{\theta^n + c^n},
\end{equation}
where $\beta_0$ denotes the maximum proliferation rate, $c$ represents the impact of cytokine signals released from all stem cells, $\theta$ represents the half-effective concentration of cytokines, and $n$ stands for the Hill coefficient. This cytokine signal often depends on the collective state of all stem cells within the niche under consideration, forming a feedback loop in cell growth. 

Proliferation, differentiation, senescence, and cell apoptosis (or cell death at the resting phase) are pivotal in determining the population dynamics of multicellular systems. Furthermore, cellular heterogeneity, characterized by variance in the epigenetic state of each cell, significantly influences these processes. The kinetic rates of proliferation, differentiation, and apoptosis of each cell are dependent on its epigenetic states. To represent this heterogeneity, we introduced a variable $\vec{x}$ (typically a high-dimensional vector) for the epigenetic state of a cell and $\Omega$ as the space encompassing all potential epigenetic states of the resting-phase stem cells\cite{lei2020general,Lei2020,2014PNAS..111E.880L}. Through the epigenetic state $\vec{x}\in \Omega$, let $Q(t, \vec{x})$ denote the count of cells in the resting phase at time $t$ with epigenetic state $\vec{x}$, and the total cell count is given by 
$$
Q(t) = \int_\Omega Q(t, \vec{x}) d \vec{x}.
$$
If $\xi(\vec{x})$ represents the rate of cytokine secretion by a cell with the state $\vec{x}$, the effective cytokine concentration regulating cell proliferation is expressed as  
$$
c(t) = \int_\Omega Q(t, \vec{x}) \zeta(\vec{x}) d \vec{x}.
$$ 
Moreover, the kinetic rates $\beta$, $\kappa$, $\mu$, and the duration of the proliferating phase $\tau$ are dependent on the epigenetic state $\vec{x}$. The quadruple $(\beta(c, \vec{x}), \kappa(\vec{x}), \mu(\vec{x}), \tau(\vec{x}))$ constitutes the \textit{kinetotype} of a cell\cite{lei2020general}.

During cell division, a single mother cell divides into two daughter cells. However, these daughter cells may not share the same epigenetic state as the mother cell, leading to cell plasticity in cellular regulation. By introducing an inheritance function (also known as the transition function) $p(\vec{x}, \vec{y})$ defined by the conditional probability density \eqref{eq:29}, the evolution equation \eqref{4} for the cell population $Q(t, \vec{x})$ can be derived from the G0 cell cycle model as follows\cite{lei2020general}:
\begin{equation}
	\label{4'}
	\left\{
	\begin{aligned}
		\frac{\partial Q(t, \vec{x})}{\partial t} =&
		-Q(t,\vec{x})(\beta(c(t), \vec{x})+\kappa(\vec{x}))\\
		&+2\int _{\Omega} \beta (c(t-\tau(\vec{y})), \vec{y}) Q(t-\tau( \vec{y}), \vec{y})e^{-\mu( \vec{y})\tau( \vec{y})}p(\vec{x}, \vec{y}) d \vec{y},\\
		c(t)=&\int_{\Omega} Q(t, \vec{x})\zeta(\vec{x})d \vec{x}.\\
	\end{aligned}
	\right.
\end{equation}
 	
Equation \eqref{4'} provides a comprehensive mathematical framework for modeling the dynamics of heterogeneous stem cell regeneration with epigenetic transitions.	

\subsection{Gene regulation network}

\subsubsection{Ordinary differential equation model}

\begin{figure}[htbp]
	\centering
	\includegraphics[width=6cm]{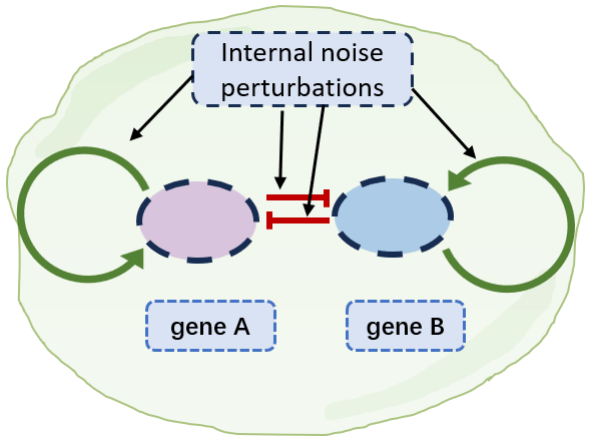}
	\caption{\textbf{Schematic diagram depicting the gene regulatory network}.  Green arrows indicate self-activating interactions of genes, and red bars indicate repressive interactions between genes. }
	\label{fig:1}
\end{figure}

We introduced gene regulation networks to establish connections between the epigenetic state and gene expressions within individual cells. For simplicity, we considered a basic gene network comprising two master transcription factors (TFs), namely gene A and gene B. These factors exhibit self-activation and mutual repression, as illustrated in Figure \ref{fig:1}. Let $X_1$ and $X_2$ denote the expression levels (protein concentration) of genes A and B, respectively. While we focus on the master genes A and B, the gene expression levels $\vec{X} = (X_1, X_2)$ define the epigenetic state of a cell.

The kinetics of gene expression within one cell cycle is modeled using the following ordinary differential equations (ODEs):
\begin{equation}
\label{6}
\left\{
\begin{aligned}
&\frac{d X_{1}}{dt}&=\quad g_{1}\left(\frac{1+\lambda_{21}(\frac{X_2}{h_{21}})^{n_{21}}}{1 + (\frac{X_{2}}{h_{21}})^{n_{21}}}\right) \left(\frac{1+\lambda_{11} (\frac{X_{1}}{h_{11}})^{n_{11}}}{1+(\frac{X_{1}}{h_{11}})^{n_{11}}}\right)-k_{1} X_{1},\\
&\frac{d X_{2}}{dt}&=\quad g_{2}\left(\frac{1+\lambda_{12}(\frac{X_1}{h_{12}})^{n_{12}}}{1+(\frac{X_{1}}{h_{12}})^{n_{12}}}\right) \left(\frac{1+\lambda_{22} (\frac{X_{2}}{h_{22}})^{n_{22}}}{1+(\frac{X_{2}}{h_{22}})^{n_{22}}}\right)-k_{2}X_{2}.\\
\end{aligned}
\right.
\end{equation}
Here, the parameters $g_{i}$, $\lambda_{ij}$, $n_{ij}$, $h_{ij}$, and $k_{i}$ ($i, j = 1,2)$ are non-negative parameters. The parameters $g_{i}$ represent the basal protein production rates of the two genes; $\lambda_{ij}$ denote the regulatory strength from gene $j$ to gene $i$, with $\lambda_{ij} < 1$ indicating repression and $\lambda_{ij} > 1$ indicating activation; $h_{ij}$ represent the half-effective protein concentrations for the corresponding regulation; $n_{ij}$ are the Hill coefficients; $k_{i}$ denote the degradation rates of the two proteins.

Folllowing the setup in RACIPE\cite{Huang:2018fd}, we define a function $H$ as
$$
H(X; h, n, \lambda) = \dfrac{1 + \lambda (\frac{X}{h})^{n}}{1 + (\frac{X}{h})^{n}},
$$ 
and equation \eqref{6} can be expressed as
\begin{equation}
\label{6'}
\dfrac{d X_i}{d t} = g_i \prod_{j=1}^2 H(X_j; h_{ji}, n_{ji}, \lambda_{ji}) - k_i X_i,\quad (i=1,2)
\end{equation}
for simplicity.

It is worth noting that the gene network in Figure \ref{fig:1} can be modeled using two different approaches: additive or multiplicative regulations, where the production rates from positive and negative feedbacks are either added or multiplied, respectively\cite{Jolly2015a,Xu2014}. This study adopts the multiplicative modeling approach, consistent with the model described by Jolly et al. (2015)\cite{Jolly2015a}. However, the computational scheme presented in this study remains consistent with additive feedback regulations.

\subsubsection{Random perturbation to model parameters}

In equation \eqref{6'}, the production rates $g_i$, effective concentrations $ h_{ij}$, and degradation rates $k_i$ may experience random fluctuations due to the extrinsic noise in the cellular environment. To incorporate the effects of these random perturbations, we introduced stochastic fluctuations to each parameter $a$ ($a = g_i, h_{ij}$, or $k_i$)  as follows: 
\begin{equation}
\label{7}
a  = \bar{a} e^{\sigma \eta - \sigma^2/2},
\end{equation}
where $\sigma$ represents the intensity of the noise perturbation, and $\eta$ follows a colored noise pattern defined by the Ornstein-Uhlenbeck process:
\begin{equation}
\label{8}
d\eta  = - (\eta/\vartheta) d t + \sqrt{2/\vartheta} d W,
\end{equation}
where $W$ denoting the Wiener process and $\vartheta$ represents the relaxation coefficient. 

Unlike the conventional Langevin stochastic differential equation approach, the random perturbation $\eta$ in \eqref{7} is expressed exponentially. This form has been previously utilized in research studies to simulate extrinsic noise perturbations in gene expression\cite{huang2018cell,huang2023dynamics,Lei2009}. The exponential form \eqref{7} avoids potential issues with negative parameter values that may arise with the Langevin approach $a \to a + \sigma \eta$. In biochemical reactions, a chemical rate $a$ is often associated with a chemical potential $\mu$ through $a = c e^{-\beta \mu}$. Consequently, a perturbation to the chemical potential $\mu$ naturally translates into an exponential form perturbation in the chemical reaction rate. Moreover, gene expression rates have been observed to follow a log-normal distribution rather than a normal distribution\cite{Austin2006}. Thus, formulating the random perturbation by multiplying the log-normal distribution random number is appropriate. Here, the random perturbation is defined by an Ornstein-Uhlenbeck process. For a more detailed discussion, please refer to Lei (2011)\cite{Lei:2011dd} and Huang et al. (2024)\cite{huang2023dynamics}.   

It is important to note that \eqref{7}-\eqref{8} provide a general formulation for noise perturbation to a parameter. However, the perturbations $\eta$ to different parameters $g_i$, $h_{ij}$ and $k_i$ in equation \eqref{6'} are independent of each other. Additionally, parameters $\sigma$ and $\vartheta$ defining the noise perturbations differ for perturbations to different parameters. Thus, the ODE model \eqref{6} can be extended as the following stochastic differential equation (SDE) model:
\begin{equation}
\label{eq:10}
\left\{
\begin{aligned}
\dfrac{d X_i}{d t} & = g_i \prod_{j=1}^2 H(X_j; h_{ji}, n_{ji}, \lambda_{ji}) - k_i X_i,\quad (i=1,2)\\
g_i  &= \bar{g}_i e^{\sigma_1 \eta_1 - \sigma_1^2/2}\\
h_{ij} &= \bar{h}_{ij} e^{\sigma_2 \eta_2 - \sigma_2^2/2}\\
k_i &= \bar{k}_i e^{\sigma_3 \eta_3 - \sigma_3^2/2}\\
d\eta_1&= -(\eta_1/\vartheta_1) d t + \sqrt{2/\vartheta_1} d W_1\\
d\eta_2&= -(\eta_2/\vartheta_2) d t + \sqrt{2/\vartheta_2} d W_2\\
d\eta_3&= -(\eta_3/\vartheta_3) d t + \sqrt{2/\vartheta_3} d W_3\\
\end{aligned}
\right.
\end{equation}
	
\subsubsection{Cross-cell-cycling dynamics}

The SDE model \eqref{eq:10} describes the gene regulation dynamics within a single cell cycle. To extend these dynamics across multiple cycles and link them with the heterogeneous G0 cell cycle model \eqref{4'}, we must integrate the cell cycling process into the equation \eqref{eq:10}.

\begin{figure}[htbp]
\centering
\includegraphics[width=8cm]{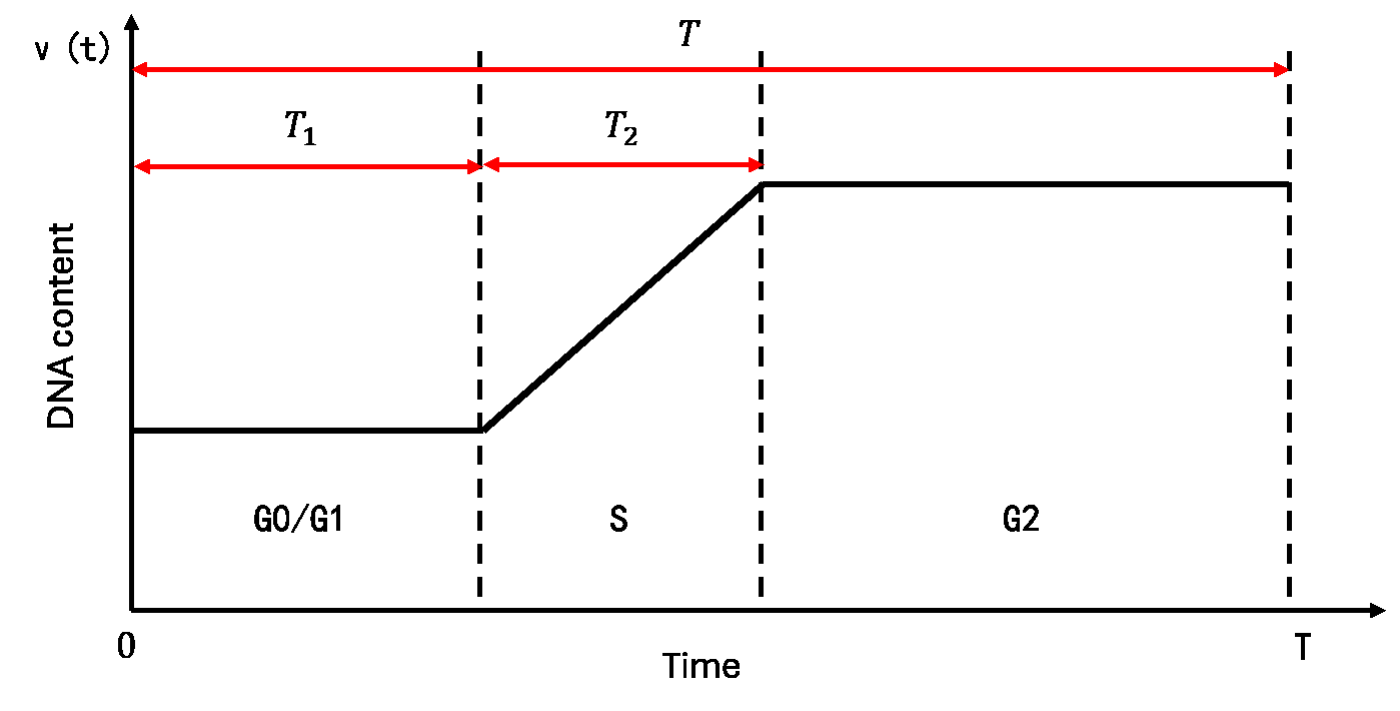}
\caption{\textbf{Illustration of a cell cycle and the cycling age.} Here $T$ denotes the duration of one cycle, $T_1$ means the duration of G0/G1 phase, and $T_2$ represents the duration of S phase.}
\label{fig:cycle}
\end{figure}

During the proliferative phase, a cell undergoes DNA replication, resulting in a change in its genetic content. Consequently, the gene expression rates vary throughout this phase. For simplicity, let $T$ represent the duration of one cycle, with the cycling age $\ell$ denoted as $0<\ell < T_1$ for the G0/G1 phase, $T_1 < \ell < T_1 + T_2$ for the S phase (DNA replication), and $T_1 + T_2 < \ell < T$ for the G2 and M phase post-DNA replication (Figure \ref{fig:cycle}). Let $\ell(t)$ represents the cycling age defining as  
\begin{equation}
\ell(t) = t - k T,\quad k T \leq t < (k+1) T, \quad (k=0,1,2,\cdots).
\end{equation}
We defined the changes in DNA content with the cell cycling age as
\begin{equation}
 	\nu(t) =
 	\left\{
 	\begin{array}{ll}
 		1, &0\leq \ell(t)<T_{1}\\
 		(1+\frac{\ell-T_{1}}{T_{2}}), &T_{1}\leq \ell(t)<(T_{1}+T_{2})\\
 		2, &(T_{1}+T_{2})\leq \ell(t) <T\\
 	\end{array}
 	\right.
\end{equation}
for simplicity. Through the DNA content $\nu(t)$, we replaced $g_i$ with $g_i \nu(t)$ and rewrote \eqref{6'} as:
\begin{equation}
\label{eq:11}
\frac{d X_{i}}{dt} = g_{i} \nu(t) \prod_{j=1}^2 H(X_j; h_{ji}, n_{ji}, \lambda_{ji}) - k_i X_i,\quad (i=1,2)
\end{equation}
when $k T \leq t < (k+1) T$.

The cell undergoes mitosis when $t = kT$. Molecules such as proteins and mRNAs redistribute stochastically to two daughter cells. Therefore, the protein concentrations $X_1$ and $X_2$ in the newborn cell should be reset post-mitosis while molecules are allocated to the two daughters. Thus, we reset the initial condition at $t = k T$ as 
\begin{equation}
\label{eq:12}
X_i(k T) = \chi_i \lim_{t \to kT^-}X_i(t)
\end{equation}
where $\chi_i$ is a beta distribution random number. 
 
\begin{figure}[htbp]
\centering
\includegraphics[width=12cm]{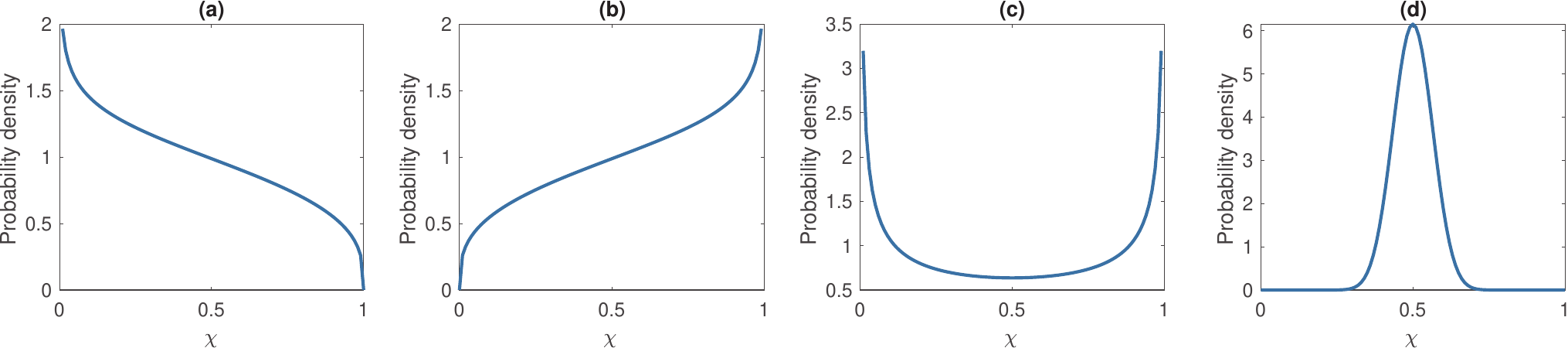}
\caption{Probability density function of the beta distribution. \textbf{(a).} $\phi=0.4$, $\eta = 2.2$. \textbf{(b).} $\phi=0.6$, $\eta=2.2$. \textbf{(c).} $\phi=0.5$, $\eta=1.0$. \textbf{(d).} $\phi=0.5$, $\eta=60$.}
\label{fig-dist}
\end{figure}

We assumed a beta distribution random number with a density function exhibiting various shapes based on shape parameters $a$ and $b$, including strictly decreasing ($a < 1, b\leq 1$), strictly increasing ($a > 1, b \leq 1$), U-shaped ($a < 1, b< 1$), or unimodal ($a>1, b>1$). We denote the beta distribution random number as $\chi_i \sim  \mathrm{Beta}(a, b)$, where shape parameters $a, b$ satisfy 
$$
\mathrm{E}(\chi_i) = \frac{a}{a + b} = \frac{1}{2}.
$$ 
We can select proper parameters to describe the situations of either symmetry or asymmetry division. In this study,  we introduced two parameters $0<\phi<1$ and $\eta>0$ so that 
\begin{equation}
\label{eq:chi2}
\mathrm{E}(\chi_i) = \phi, \mathrm{var}(\chi_i) = \dfrac{1}{1 + \eta} \phi (1-\phi),
\end{equation}
then the shape parameters $a$ and $b$ are given by
$$
a = \eta \phi, b = \eta (1-\phi).
$$
In particular, we took $\phi = \frac{1}{2}$, then $\eta > 2$  yields a unimodal distribution, corresponding to symmetry division, while $\eta<2$ gives a U-shape distribution, corresponding to asymmetry division (Figure \ref{fig-dist}).

Therefore, considering cell cycling, we integrated equations \eqref{eq:10}-\eqref{eq:12} to formulate a hybrid model as  follows (where $m =2$ represents the number of genes):
\begin{equation}
\label{eq:13}
\left\{
\begin{aligned}
\dfrac{d X_i}{d t} & = g_i \nu(t) \prod_{j=1}^m H(X_j; h_{ji}, n_{ji}, \lambda_{ji}) - k_i X_i,\quad (i=1,\cdots, m)\\
g_i  &= \bar{g}_i e^{\sigma_1 \eta_1 - \sigma_1^2/2}\\
h_{ij} &= \bar{h}_{ij} e^{\sigma_2 \eta_2 - \sigma_2^2/2}\\
k_i &= \bar{k}_i e^{\sigma_3 \eta_3 - \sigma_3^2/2}\\
d\eta_1&= -(\eta_1/\vartheta_1) d t + \sqrt{2/\vartheta_1} d W_1\\
d\eta_2&= -(\eta_2/\vartheta_2) d t + \sqrt{2/\vartheta_2} d W_2\\
d\eta_3&= -(\eta_3/\vartheta_3) d t + \sqrt{2/\vartheta_3} d W_3\\
&\qquad (k T \leq t < (k+1) T)\\
X_i(k T) &= \chi_i \lim_{t \to kT^-} X_i(t)\\
\chi_i & \sim \mathrm{Beta}(a, b).
\end{aligned}
\right.
\end{equation}

In the equation \eqref{eq:13}, the coefficient $\lambda_{ji}$ in the function $H(x_j; h_{ji}, n_{ji}, \lambda_{ji})$ represents positive regulation ($\lambda_{ji} > 1$), negative regulation ($\lambda_{ji} < 1)$, or no regulation ($\lambda_{ji} = 0$) between genes $j$ and $i$. Thus, it is easy to extend the model equation to the dynamics of a general gene regulation network with $m$ genes.

The equation \eqref{eq:13} describes the dynamics of the gene regulation network over multiple cell cycles, with the protein levels reset after cell division. 

This study does not consider the effect of cell volume change. Cell growth and volume changes are significant biological processes that play essential roles in cell-fate decisions\cite{Doncic2011,Ginzberg2015}. Moreover, biological mechanisms controlling cell growth and division are complex and not fully understood for mammalian cells\cite{Kafri2013,Cadart2018,Zatulovskiy2020}. In this study, we focused on the computational scheme connecting gene regulation network dynamics with stem cell regeneration to avoid complexity. We assumed that the volume remains unchanged during cell cycling, so cell growth does not affect the protein concentration. 

\subsubsection{Numerical implementation}
	
Implementing the model equation \eqref{eq:13} numerically includes solving the SDE model within a cell cycle and managing the process of cell division. Cell division may lead to cell splitting and the resetting of initial conditions for newborn cells. This numerical scheme can be realized using the object-oriented programming language \verb|C++|. The dynamics of gene regulatory networks within a cell cycle are modeled with the SDE model, which is numerically solved through the Runge-Kutta method. Additionally, we designed an individual-cell-based numerical scheme to monitor the dynamics of each cell within the system. Specific algorithm details are provided in the Appendix.
 	
In our numerical implementations, we focused on symmetric gene regulations, assuming $k_{1} = k_{2}$, $\lambda_{12} = \lambda_{21}$, $\lambda_{11} = \lambda_{22}$, $n_{11} = n_{12} = n_{21} = n_{22}$, and normalized the protein concentrations so that $h_{11} = h_{12} = h_{21} = h_{22}$. The parameter values were estimated based on previous studies for the similar gene network\cite{huang2023dynamics,Jolly2015a,huang2016interrogating}, and further adjustments were made to ensure that cells could exhibit various phenotypes with different gene expression levels.
		  	 	 
\subsection{Inheritance function}
	
Connecting the dynamics of the gene regulation network with the heterogeneous cell cycle model \eqref{4'} requires identifying the epigenetic state $\vec{x}$ and deriving the inheritance function $p(\vec{x}, \vec{y})$ based on the gene regulation dynamics \eqref{eq:13} across a cell cycle.  

It is important to note that while the epigenetic state $\vec{x}$ in \eqref{4'} represents a constant reflecting the state of a cell during the resting phase, it dynamically changes over time in \eqref{eq:13}, representing gene expression dynamics during a cell cycle. We need to identify the epigenetic state $\vec{x}$ in accordance with the gene expression dynamics $\vec{X}(t)$ in one cell cycle.  For consistency, we define the epigenetic state $\vec{x}$ of a cell through log normalization of gene expression $\vec{X}(t)$ as
\begin{equation}
\vec{x} = \log(\vec{X}(t) + 1)\vert_{\ell(t) = \ell^*}
\end{equation}
with the time $t$ corresponding to a specific cell cycling age $\ell^*$. In our simulations, we took $T = 50$, and $T_1 = 25, T_2 = 8$. We considered the state at $\ell^* = T_1$, which is the time point before DNA replication, as the cycling age to define the epigenetic state of a cell during a cell cycle.   

To obtain the inheritance function from simulation results, we numerically solved the cross-cell-cycle stochastic differential equation \eqref{eq:13} and tracked the gene expression $\vec{X}(t)$ of individual cells over cell divisions. Since \eqref{eq:13} describes the gene expression dynamics of a single cell over multiple cycles, two daughter cells were recorded when a cell divided. Thus, we repeated the numerical scheme to obtain an ensemble of multiple cells, with each simulation starting with a randomly selected initial condition. For each cell under simulation, record the state $\vec{X} = \vec{X}(t)|_{\ell = T_1}$ so that $\vec{x} = \log (\vec{X} + 1)$ represents the epigenetic state of the cell. This process yields a dataset $D = \{\vec{x}_k, \vec{y}_k\}$, where $\vec{y}_k$ represents the state of the $k$'th mother cell and $\vec{x}_k$ the state of one daughter cell of the $k$'th mother cell. Through dataset $D$, we obtain the inheritance function $p(\vec{x}, \vec{y})$ via the conditional probability density defined by \eqref{eq:29}.

For the probability density function $p(\vec{x}, \vec{y})$, we assumed that the inheritances of gene A and gene B were independent of each other, yielding 
$$
p(\vec{x}, \vec{y}) = \prod_{i=1}^2 p_i(x_i, \vec{y}),
$$
where $p_i(x_i, \vec{y})$ denotes the conditional density function of $x_i$ given the state of the mother cell $\vec{y} = (y_1, y_2)$. 

In our study, we presumed the mixed conditional gamma distribution for epigenetic states so that
\begin{equation}
\label{eq:16}
p_i(x_i, \vec{y}) = \sum_{j=1}^k \alpha_{i,j}\mathrm{Gamma}(x_i; a_{i,j}, b_{i,j}),\quad (i=1, 2),
\end{equation}
where $k$ is the number of independent distributions, $\alpha_{i,j}$ are combination coefficients, and $\mathrm{Gamma}(x; a, b)$ represents the probability function of gamma distribution with shape parameters $a$ and $b$, expressed as
\begin{equation}
\label{eq:17}
 \mathrm{Gamma}(x; a, b)  =  \dfrac{x^{a-1} e^{-(x/b)}}{b^{a} \Gamma(a)}, \quad \Gamma(z) = \int_0^\infty e^{-x} x^{z-1} d x, \quad z > 0.
 \end{equation}
The combination coefficients $\alpha_{i,j}$ satisfy
$$
\sum_{j=1}^k \alpha_{i,j} = 1,\quad \alpha_{i,j} \geq 0.
$$
In \eqref{eq:16}, the shape parameters $a_{i,j}$ and $b_{i,j}$ depend on the state $\vec{y}$ of the mother cell. 

To determine the shape parameters, we consider that, given the probability density function \eqref{eq:17}, the conditional mean and variance are
$$
\mathrm{E}(x | \vec{y}) = a(\vec{y}) b(\vec{y}),\quad \mathrm{var}(x | \vec{y}) = a(\vec{y}) b(\vec{y})^2. 
$$
Thus, while we assume the conditional mean and variance as
\begin{equation}
\label{eq:19}
\mathrm{E}(x | \vec{y}) = \psi(\vec{y}),\quad \mathrm{var}(x | \vec{y}) = \dfrac{\psi(\vec{y})^2}{\zeta(\vec{y})},
\end{equation}
we have
\begin{equation}
\label{eq:20}
a(\vec{y}) = \zeta(\vec{y}),\quad b(\vec{y}) = \psi(\vec{y})/\zeta(\vec{y}). 
\end{equation}

Thus, to obtain the inheritance function $p(\vec{x}, \vec{y})$, we only need to derive, for each component $x_i$, the functions $\psi_{i,j}(\vec{y})$ and $\zeta_{i,j}(\vec{y})$ through the conditional mean and variance from the simulation data. The detailed procedure for obtaining the inheritance function is provided in Section \ref{sec:3.2}.
	
\section{Results}
\subsection{Identification of cell types through epigenetic states}
 	
\begin{figure}[htbp]
 	\centering
 	\includegraphics[width=10cm]{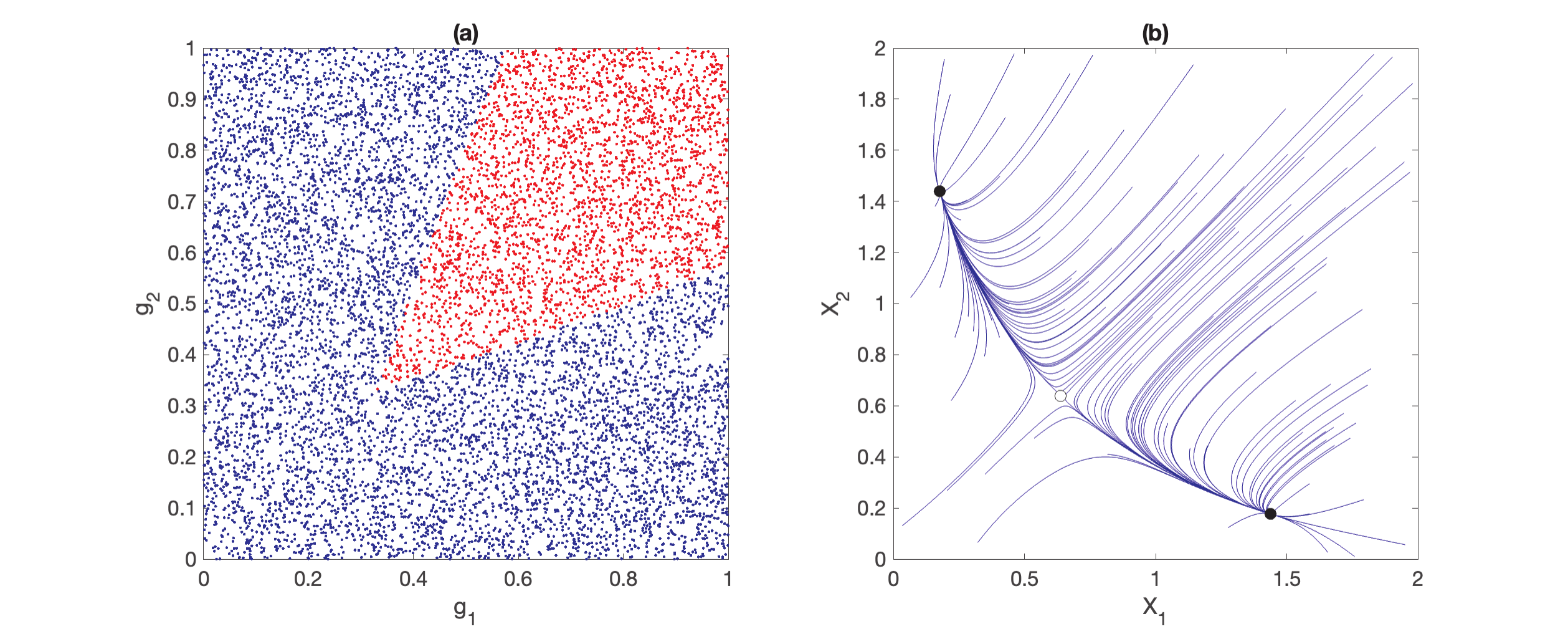}
 	\caption{\textbf{Bifurcation analysis of gene expression kinetics}. \textbf{(a)}. Dependence of the number of steady states of the ODE model \eqref{6} with randomly selected parameters $(g_1, g_2)$. Blue dots represent 1 steady state, while red dots represent 3 steady states. \textbf{(b)}. Trajectories for the ODE model \eqref{6} with randomly selected initial conditions. Solid dots marked the two stable steady states, and the hollow dot represents the unstable steady state. Parameters are $\lambda_{21} = \lambda_{12} = 0.1$, $\lambda_{11}=\lambda_{22}=5.0$, $k_1 = k_2 = 1.0$, $h_{ij} = 1$, $n_{ij} = 2$, and $(g_1, g_2) = (0.4, 0.4)$ in (b). The parameters for time units have been normalized in our study.}
 	\label{fig:3}
\end{figure}

To quantitatively define cell phenotypes, we conducted a bifurcation analysis based on the ODE model \eqref{6}. We varied the production rates $g_1$ and $g_2$ while keeping other parameters fixed and examined the dependence of steady states on these parameter values. Figure \ref{fig:3}a illustrates the number of steady states with randomly varying parameters $(g_1, g_2)$. When $g_1$ and $g_2$ are large, three steady states emergy (depicted by red dots in Figure \ref{fig:3}a). Additionally, at $(g_1, g_2) = (0.4, 0.4)$, solving the ODE \eqref{6} with varying initial conditions results in the convergence of solutions $(x_1(t), x_2(t))$ to one of the two stable steady states, as indicated by the black dots in Figure \ref{fig:3}b. These stable steady states exhibit expression patterns of either $(+, -)$ or $(-, +)$ for the maker genes A and B, defining two distinct cell phenotypes. 

The bifurcation analysis reveals bistable steady states in gene network dynamics, corresponding to different cell types defined by marker gene expression patterns. To further explore how external noise perturbation and cell divisions may shape the heterogeneous within a cell clone, we examined the dynamics of the hybrid model \eqref{eq:13} under various noise strength $\sigma$ (assuming $\sigma_1 = \sigma_2 = \sigma_3 = \sigma$) and cell division scenarios.  

\begin{figure}[htbp]
 	\centering
 	\includegraphics[width=10cm]{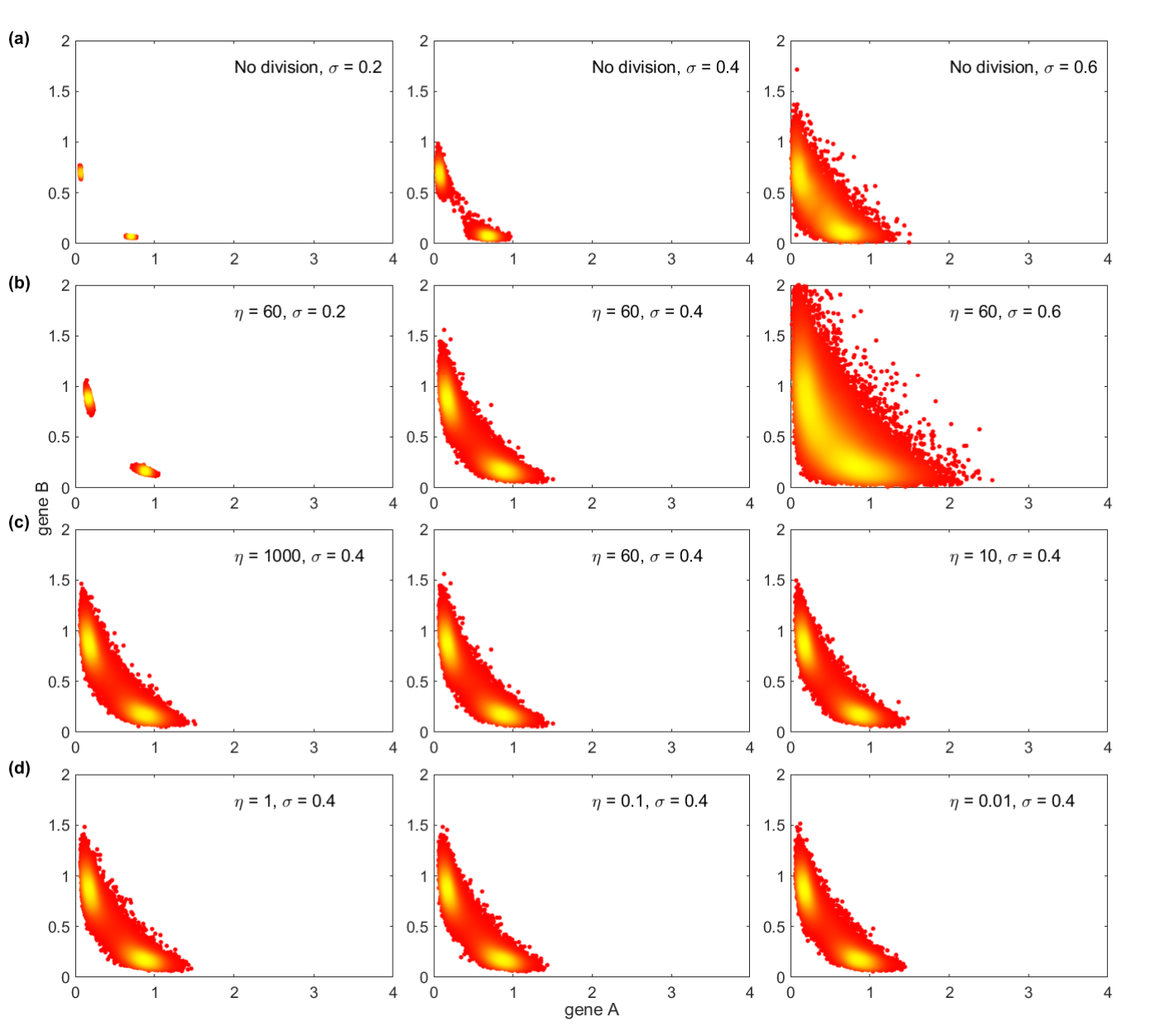}
 	\caption{\textbf{Distribution of the epigenetic state of cells under different model assumptions}. \textbf{(a).} Epigenetic states of cells at $t = 50$ (one cell cycle) without cell division. The cell states were obtained by solving the SDE model \eqref{eq:10} with initial conditions randomly distributed over $0< X_i(0)< 2$. \textbf{(b).}  Epigenetic state of cells derived from a single cell after 15 divisions. The cell states were obtained by solving the hybrid model \eqref{eq:13}, with initially a single cell having random gene expression states $0< X_i(0)< 2$. \textbf{(c).} Same as (b) with symmetric division ($\eta > 2$) and different distributions for the parameter $\chi_i$. \textbf{(d).} Same as (b) with asymmetric division ($\eta < 2$) and different distributions for the parameter $\chi_i$. Here, the epigenetic states are represented by $\vec{x} = \log(\vec{X}(t) + 1)|_{\ell(t) = T_1}$. In (a), no cell division was considered, and different external perturbation strengths $\sigma$ were applied. In (b)-(d), cell division was considered, with $\phi = 0.5$ and different parameters $\eta$ for the distribution of $\chi_i$ (refer to \eqref{eq:chi2}). The values of $\sigma$ and $\eta$ are shown in the figure; the parameters $\vartheta_1 = \vartheta_2 = \vartheta_3 = 0.3$; the conditional perspective $\phi$ in \eqref{eq:chi2} was set as $\phi =0.5$; the cell cycling parameters were $T=50, T_1 = 25, T_2 = 8$; other parameters were the same as in Figure \ref{fig:3}.}
	\label{fig:4}
\end{figure}

First, we excluded cell cycling and focused solely on gene expression dynamics with external noise perturbation. To achieve this, we solved the SDE model \eqref{eq:10} with randomly selected initial conditions $0< X_i(0) < 2$. Figure \ref{fig:4}a illustrates the distributions of the epigenetic state of cells at one cycle ($T=50$) with varying noise strength $\sigma$. Results indicate distinct cell phenotypes when the external noise is small ($\sigma =0.2$ or $0.4$); however, the two cell types merge as the noise strength $\sigma$ increases.
 	 	
Next, we explored the influence of cell division using the hybrid model \eqref{eq:13}. The effect of cell division is depicted by the redistribution of molecules through \eqref{eq:12}, with parameters $\phi$ and $\eta$ defining the distribution of the coefficient $\chi_i$. We fixed $(\phi, \eta) = (0.5, 60)$ (Figure \ref{fig-dist}d) and varied the noise strength $\sigma$ to solve \eqref{eq:13}. Here, we initiated the system with one cell and simulated model dynamics for 15 cycles, resulting in $2^{15}$ cells at the $15$th cycle, all originating from a single cell. Figure \ref{fig:4}b displays the epigenetic states of all cells, showing a similar pattern to Figure \ref{fig:4}a, generated from multiple cells with different initial conditions. This suggests that a single cell can give rise to a heterogenous cell clone through cell division.    
 	 	
Furthermore, to investigate the effects of different assumptions in cell division, we varied the parameter $\eta$ in defining the distribution of the coefficient $\chi_i$. Fixing $\sigma = 0.4$, we adjusted $\eta$ to reflect symmetry division ($10$ to $1000$), resulting in a decreased variance of $\chi_i$. The distribution of epigenetic states in Figure \ref{fig:4} appears independent of the parameter $\eta$. Similarly, varying $\eta$ to reflect asymmetric division (ranging from $1$ to $0.01$) yields consistent distributions of epigenetic states at the $15$th cycle, as shown in Figure \ref{fig:4}d. These results suggest that the distribution of the epigenetic states of a cell clone remains insensitive to the variance of $\chi_i$ in cell division. 

For subsequent discussions below, we focused on the method of obtaining the inheritance function through simulation data, with fixed values of $\phi = 0.5, \eta = 60$, and $\sigma = 0.4$.
 
\subsection{Data-driven inheritance function}
 \label{sec:3.2}
 
In Figure \ref{fig:4}, we have demonstrated the variability of epigenetic states in individual cells originating from a single cell. The variability reflects the stochastic changes in cell states during cell division. To derive the inheritance function $p(\vec{x}, \vec{y})$ based on the gene regulation dynamics \eqref{eq:13}, we tracked the process of cell division through numerical simulations. We recorded the state of the mother cells ($\vec{y}$) and the daughter cells ($\vec{x}$) following each cell division event. In our simulations, we initiated $10^6$ cells. For each cell, we simulated the model for 3 cell cycles, considering all cells at the second cycle as mother cells and denoting their epigenetic states as $\vec{y}$. For each mother cell with state $\vec{y}$, we paired it with the epigenetic state $\vec{x}$ of the corresponding daughter cell at the third cycle, forming a pair $[\vec{x}, \vec{y}] = [(x_1, x_2), (y_1, y_2)]$. Here, $x_1$ and $y_1$ represent the state associated with gene A, while $x_2$ and $y_2$ represent the state associated with gene B.

\begin{figure}[htbp]
 	\centering
 	\includegraphics[width=12cm]{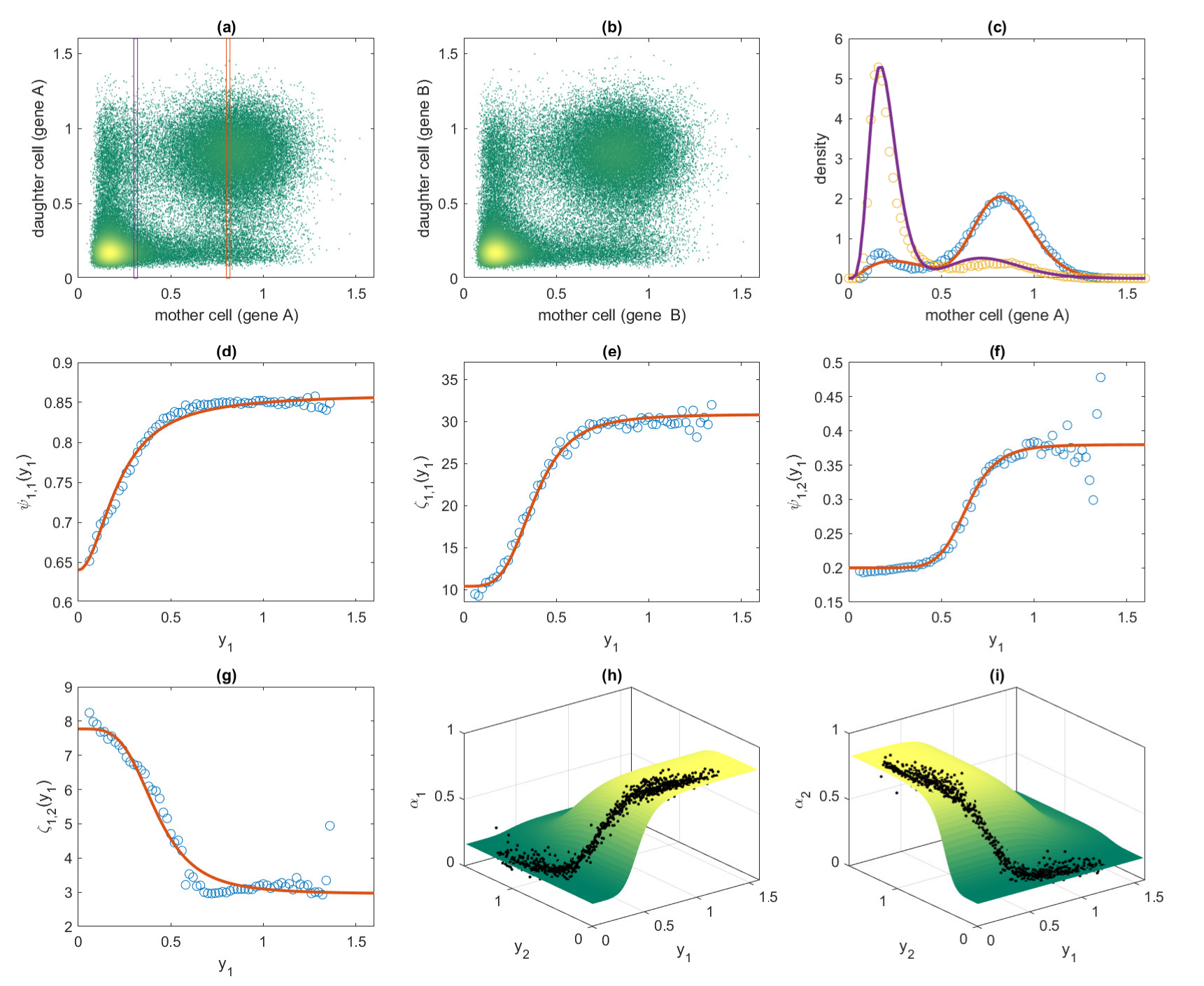}
 	\caption{\textbf{Data-driven inheritance function}. \textbf{(a)}. Scatter plot of daughter cell (gene A) versus mother cell (gene B) ($x_1$ versus $y_1$). Each point represents a cell, with lighter colors indicating higher density. Two vertical strips represent the sampling of data points with $0.3 \leq y_1<0.31$ and $0.8 \leq y_1 < 0.81$, respectively. \textbf{(b)}. Scatter plot of daughter cell (gene B) versus mother cell (gene B) ($x_2$ versus $y_2$). \textbf{(c)}. Probability density of daughter cells (gene A) given the state of the mother cell (gene A). Dots were obtained from data analysis, and solid lines represent the functions of a combined gamma distribution. The densities were obtained from data in the two strips in (a). \textbf{(d)}. The function $\psi_{1,1}(y_1)$. \textbf{(e)}. The function $\zeta_{1,1}(y_1)$. \textbf{(f)}. The function $\psi_{1,2}(y_1)$. \textbf{(g)}. The function $\zeta_{1,2}(y_1)$. \textbf{(h)}. The function $\alpha_1(y_1, y_2)$. \textbf{(i)}. The function $\alpha_2(y_1, y_2)$. In (d)-(i), dots were obtained from data analysis, while curves or surfaces were derived from mathematical formulations.}
 	\label{fig:7}
\end{figure}

Figure \ref{fig:7}a and \ref{fig:7}b present scatter plots illustrating the epigenetic states of genes A and B in mother cells and their daughter cells. These plots reveal distinct cell types: mother cells with lower expression in gene A (or gene B) typically produce daughter cells with similarly lower expression levels in the same gene, while those with higher expression tend to generate daughter cells with elevated expression levels in the same gene. However, we also observe instances of cell type switches during cell divisions, where some daughter cells exhibit gene expression patterns differing from those of their mother cells. 
 
To establish the mathematical formulation of the inheritance function, we utilized gene A as an example to illustrate the computational method. We identified all mother cells with a specific expression level (as delineated by the two strips in Figure \ref{fig:7}a) and examined the probability density of the expression levels of all daughter cells. Figure \ref{fig:7}c illustrates the resulting conditional probability densities with the epigenetic state of mother cells $y_1=0.3$ and $y_1 = 0.8$, respectively. These density functions exhibit characteristics akin to a combination of two unimodal distributions (solid lines in Figure \ref{fig:7}c). 

Drawsing insights from Figure \ref{fig:7}c, we modeled the probability density $p_1(x_1, \vec{y})$ with a combination of two gamma distributions:
\begin{equation}
\label{eq:22}
p_1(x_1, \vec{y}) = \alpha_{1} \times \mathrm{Gamma}(x_1; a_{1, 1}, b_{1, 1}) + (1-\alpha_1) \times \mathrm{Gamma}(x_1; a_{1, 2}, b_{1, 2}).
\end{equation}
Here, the shape coefficients $a_{1,j}$ and $b_{1,j}$ depend on the state of the mother cell $\vec{y}$ through the functions $\psi_{1,j}(\vec{y}|)$ and $\zeta_{1,j}(\vec{y})$, which are associated with the conditional mean and variance $\mathrm{E}(x_1 | \vec{y})$ and $\mathrm{var}(x_1 | \vec{y})$ in accordance with \eqref{eq:19}. Additionally, the combination coefficients $\alpha_{1}$ may vary based on the state of the mother cell $\vec{y}$. 

To determine the coefficients $\alpha_{1}$, $a_{1,j}$ and $b_{1,j}$, we initially presumed their dependence solely on the component $y_1$. From Figure \ref{fig:7}a, we partitioned the data into multiple bins based on the $y_1$ value. Subsequently, we segregated the $x_1$ values in each bin into two subgroups and calculated the mean and variance in each subgroup, thus deriving the conditional means $E(x_1 | y_1)$ and variances $\mathrm{var}(x_1 | y_1)$ for the two gamma distributions $\mathrm{Gamma}(x_1; a_{1,j}, b_{1,j})$, $j=1,2$. This process yielded the functions $\psi_{1,j}(y_1)$ and $\zeta_{1,j}(y_1)$ as Hill-type functions following \eqref{eq:19} (Figure \ref{fig:7}d-g): 
\begin{equation}
\label{eq:23}
\left\{
\begin{aligned}
\psi_{1,1}(y_1) &= 0.64 + 0.22\times\dfrac{ y_1^2}{0.22^2 + y_1^2},\\
\zeta_{1,1}(y_1) &= 10.38 + 20.46\times \dfrac{y_1^4}{0.38^4 + y_1^4}, \\
\psi_{1,2}(y_1) &= 0.20 + 0.18 \times\dfrac{y_1^8}{0.64^8 + y_1^8},\\
\zeta_{1,2}(y_1) &= 2.95 + 0.15\times\dfrac{1}{0.42^4 + y_1^4}.
\end{aligned}
\right.
\end{equation}
The shape parameters were formulated as
\begin{equation}
a_{1,j}(y_1) = \psi_{1,j}(y_1), b_{1,j}(y_1) =\psi_{1,j}(y_1)/\zeta_{1,j}(y_1).
\end{equation}

Subsequently, we assumed that the combination coefficient $\alpha_1$ depends on both $y_1$ and $y_2$. Consequently, we partitioned the $(y_1, y_2)$ phase plane into multiple bins and gathered the daughter cell epigenetic states $x_1$ in each bin. We then employed the expectation-maximization (EM) algorithm, as per \eqref{eq:22}, to determine the coefficients $\alpha_{1,j}$ for each bin. The results, illustrated in Figure \ref{fig:7}h, closely align with a Hill-type function:
\begin{equation}
\alpha_1(\vec{y}) = 0.16 + \dfrac{y_1^6}{0.46^6 + y_1^6}\times \dfrac{1}{1.07^6+y_2^6}.
\end{equation}
Thus, the function $\alpha_1(\vec{y})$ and the shape parameters $a_{1,j}(y_1), b_{1,j}(y_1)$ collectively define the inheritance function $p_1(x_1; \vec{y})$ per  \eqref{eq:22}.

Similarly, we derive the inheritance function $p_2(x_2; \vec{y})$ for gene B:
\begin{equation}
\label{eq:26}
p_2(x_2; \vec{y}) = \alpha_{2}(\vec{y}) \times \mathrm{Gamma}(x_2; a_{2, 1}, b_{2, 1}) + (1-\alpha_2(\vec{y})) \times \mathrm{Gamma}(x_2; a_{2, 2}, b_{2, 2}).
\end{equation}
Remarkably, owing to the symmetry between gene A and gene B, simulation data exhibit the same forms of $\psi_{2,j}(y_2)$ and $\zeta_{2,j}(y_2)$ as the functions $\psi_{1,j}(y_1)$ and $\zeta_{1,j}(y_1)$, respectively. Consequently: 
\begin{equation}
\left\{
\begin{aligned}
\psi_{1,1}(y_2) &= 0.64 + 0.22\times\dfrac{ y_2^2}{0.22^2 + y_2^2},\\
\zeta_{1,1}(y_2) &= 10.38 + 20.46\times \dfrac{y_2^4}{0.38^4 + y_2^4}, \\
\psi_{1,2}(y_2) &= 0.20 + 0.18 \times\dfrac{y_2^8}{0.64^8 + y_2^8},\\
\zeta_{1,2}(y_2) &= 2.95 + 0.15\times\dfrac{1}{0.42^4 + y_2^4},
\end{aligned}
\right.
\end{equation}
and
\begin{equation}
a_{2,j}(y_2) = \psi_{2,j}(y_2), b_{2,j}(y_2) =\psi_{2,j}(y_2)/\zeta_{2,j}(y_2).
\end{equation}
Likewise, $\alpha_2(\vec{y})$ assumes the same form as $\alpha_1(\vec{y})$, thus:
\begin{equation}
\alpha_2(\vec{y}) = 0.16 +  \dfrac{1}{1.07^6+y_1^6}\times \dfrac{y_2^6}{0.46^6 + y_2^6}.
\end{equation}
The function $\alpha_2(\vec{y})$ is shown in Figure \ref{fig:7}i.

Finally, the inheritance function is expressed as:
\begin{equation}
\label{eq:29}
p(\vec{x}; \vec{y}) = p_1(x_1; \vec{y}) \times p_2(x_2; \vec{y}).
\end{equation}
The above procedure delineates the processes of acquiring the inheritance function in \eqref{4'} from simulation data based on the gene regulation network model in \eqref{eq:13}. Validation of the inheritance function \eqref{eq:29} is presented in section \ref{sec:3.3}.

The gamma distribution density functions $\mathrm{Gamma}(x_i; a_{i,j}(y_i), b_{i,j}(y_i))$ defined by \eqref{eq:22}, \eqref{eq:26}, and \eqref{eq:29} describe the inheritance of daughter cells from the mother cells for the same gene, while the combination coefficients $\alpha_1(\vec{y})$ and $\alpha_2(\vec{y})$  elucidate the property of cell type transition regulated by the gene regulation network. Notably, $\alpha_1(\vec{y})$ increases with $y_1$ and decreases with $y_2$, signifying the activation of gene A by itself and the inhibition of gene B to gene A. A similar pattern is observed in $\alpha_2(\vec{y})$. Thus, the gene regulation network structure influences the inheritance function through its qualitative dependence on the state of mother cells.

\subsection{Validation of the inheritance function}
\label{sec:3.3}
 	
To verify the data-driven inheritance function, we compared simulation results obtained from the heterogeneous G0 cell cycle model \eqref{4'} and the gene regulation network model  \eqref{eq:13}. 

Initially, we utilized the cell cycle model \eqref{4'} to reproduce the process of cell clone generation from a single cell and compared our findings with those in Figure \ref{fig:4}. To achieve this, we defined the kinetotype in \eqref{4'} as follows:
\begin{equation}
\label{eq:31}
\beta = 1/T_1, \kappa = \mu = 0, \tau = T - T_1.
\end{equation}
Given that solving the high-dimensional integral in \eqref{4'} directly is computationally expensive, we applied an individual-cell-based simulation approach, adhering to the scheme outlined in equation \eqref{4'}. 

To replicate the process illustrated in Figure \ref{fig:7}, we initialized the system with $10^6$ cells, emulating the setup in Figure \ref{fig:7}, and performed individual-cell-based simulations to model the cell division process. During cell division, each mother cell generated two daughter cells, with the epigenetic state of each daughter cell determined by the inheritance function $p(\vec{x}, \vec{y})$ defined in \eqref{eq:29} (see the Appendix for the detailed numerical scheme). Similar to the methodology in Figure \ref{fig:7}, we conducted simulations over 3 cycles and collected the cells from the $2$nd and the $3$rd cycles for data analysis. 

\begin{figure}[htbp]
 	\centering
 	\includegraphics[width=10cm]{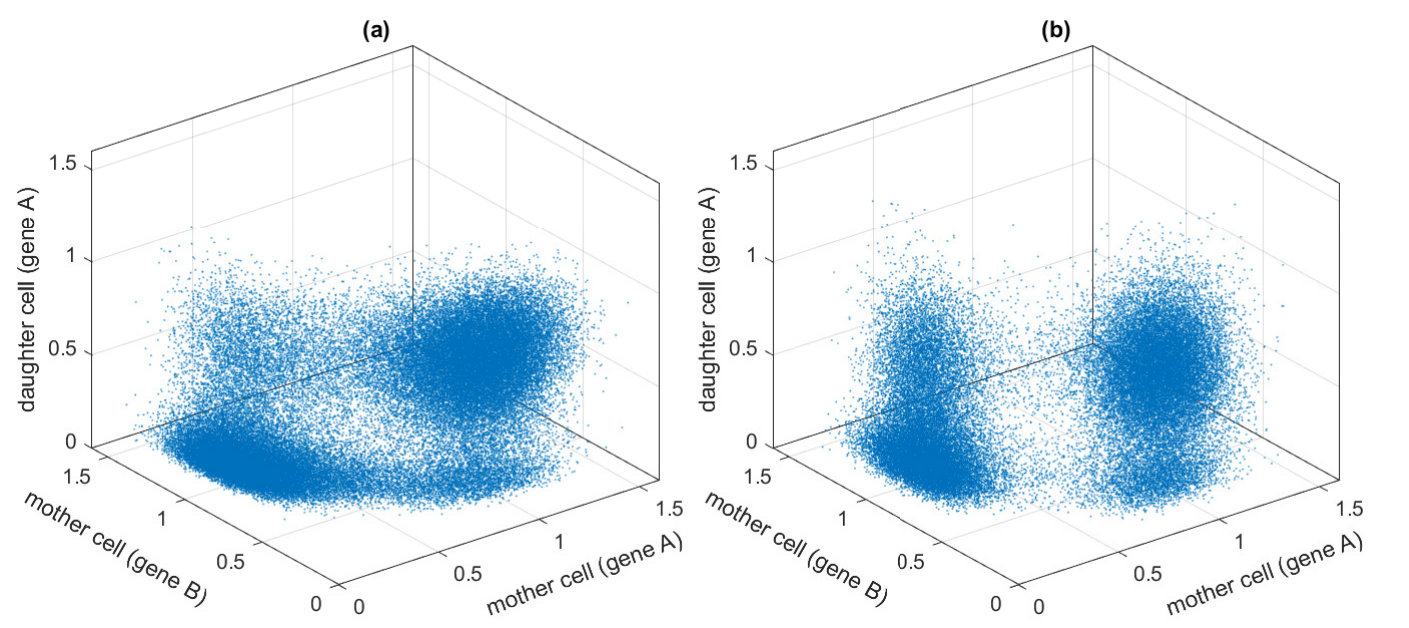}
 	\caption{\textbf{Comparison between the gene regulation network model and the G0 cell cycle model}. \textbf{(a)}. Scatter plot of daughter cell epigenetic state (gene A) versus mother cell epigenetic state (gene A and B) following the gene regulation network model \eqref{eq:13}. \textbf{(b)}. Scatter plot of daughter cell epigenetic state (gene A) versus mother cell epigenetic state (gene A and B) following the G0 cell cycle model \eqref{4'} with inheritance function given by \eqref{eq:29}. Parameters were identical to those in Figure \ref{fig:7}, with cell cycling parameters given by \eqref{eq:31} in G0 cell cycle model.}
 	\label{fig:8}
\end{figure}

Figure \ref{fig:8} presents the data points of daughter cell epigenetic states against mother cell epigenetic states, obtained from both the gene regulation network model \eqref{eq:13} and the G0 cell cycle model \eqref{4'} with the inheritance function \eqref{eq:29}. A comparison of Figure \ref{fig:8}a with Figure \ref{fig:8}b demonstrates that the data-driven inheritance function \eqref{eq:29} effectively reproduces the distribution of the cross-cell cycling epigenetic states. 

Next, we applied the G0 cell cycle model to investigate the biological process of heterogeneous cell growth with additional regulations in proliferation and differentiation. For this purpose, we defined the kinetotype as:
$$
\kappa = 0.009, \mu = 0.0007, \tau = T - T_1,  
$$ 
and
$$
\beta(c, \vec{x}) = \beta_0 \frac{\theta}{\theta + c}, c=\int_\Omega Q(t, \vec{x}) d \vec{x}, 
$$
with $\beta_0 = 0.2$ and $\theta = 10^6$. Employing identical assumptions, we simulated the cell growth process following the G0 cell cycle model \eqref{4'} with the inheritance function \eqref{eq:29}. In our simulations, we initialized $10^4$ cells, randomly assigning epigenetic states to each cell. We employed individual-cell-based simulation to model cell growth up to $t = 2500$ (about $50$ cycles).

\begin{figure}[htbp]
\centering
\includegraphics[width=10cm]{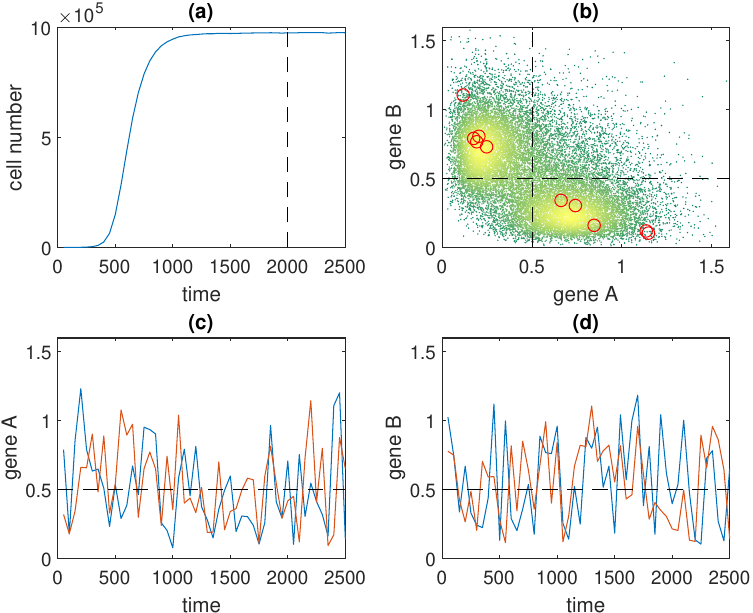}
\caption{\textbf{Heterogeneous cell growth.} \textbf{(a)}. Dynamics of cell number evolution. \textbf{(b)}. Scatter plot of the epigenetic state of cells at $t = 2000$ (dashed line in (a)). \textbf{(c}. Evolution of gene A epigenetic state of two cells during the cell growth process. \textbf{(d)}. Evolution of gene B epigenetic state of two cells during the cell growth process.  }
\label{fig:9}
\end{figure}

In this scenario, we introduced nonzero differentiation and apoptosis rates, assuming that the proliferation rate decreases with the total cell number. Consequently, the cell number would eventually reach an equilibrium state following an extended period of cell growth (Figure \ref{fig:9}a). Figure \ref{fig:9}b displays the distribution of the epigenetic state of cells at $t = 2000$ (marked by the dashed line in Figure \ref{fig:9}a). Notably, two subtypes of cells are evident, with their epigenetic state distributions closely resembling those obtained from the gene regulation network model in Figure \ref{fig:4}. We further examined the dynamics of individual cell-type transitions. Figure \ref{fig:9}c-d illustrates the evolution dynamics of gene A and gene B while tracking a single cell during the cell growth process. Our simulations demonstrate that cells can transition between the two subtypes as the cell growth process unfolds. These results underscore the capability of the G0 cell cycle model \eqref{4'} to simulate the long-term dynamics of cell growth with heterogeneity.   
 	
\section{Discussions}

Our study investigated the inheritance of epigenetic states in cells using a combination of computational modeling and data analysis. We developed a data-driven inheritance function based on the gene regulation dynamics model and validated it using simulations from both a gene regulation network model and a heterogenous G0 cell cycle model. Our results demonstrate that the data-driven inheritance function effectively captures the distribution of epigenetic states across cell generations. Furthermore, we applied the G0 cell cycle model to simulate the long-term dynamics of cell growth with heterogeneity, showcasing its capability to replicate the behavior observed in the gene regulation network model.

The findings of our study shed light on the mechanisms underlying the inheritance of epigenetic states in cells. By combining computational modeling with data analysis, we were able to derive a robust inheritance function that accurately predicts the distribution of epigenetic states in daughter cells based on their mother cell's states. This suggests that epigenetic inheritance is governed by complex regulatory processes embedded within the gene regulation network.

Moreover, our simulations using the G0 cell cycle model revealed intriguing insights into the dynamics of cell growth with heterogeneity. The model successfully reproduced the emergence of distinct cell subtypes and their transitions over time, indicating that the heterogeneity observed in experimental data can be attributed to underlying regulatory mechanisms within the cell cycle.

In conclusion, our study provides valuable insights into the inheritance of epigenetic states and the dynamics of cell growth with heterogeneity. By developing and validating a data-driven inheritance function, we have contributed to a deeper understanding of the regulatory processes governing cell behavior. Moving forward, our findings could inform further research into the molecular mechanisms underlying epigenetic inheritance and their implications for cellular function and disease.

\section*{Acknowledgments}
This work was funded by the National Natural Science Foundation of China (NSFC 12331018).
	
\section*{Appendix}
\subsection*{A. Numerical scheme for the hybrid kinetic modeling}

We applied the numerical scheme to solve the hybrid model \eqref{eq:13} by combining the numerical method of stochastic differential equation and the random process of cell division. During numerical simulation, the cell number is changed over time due to cell division. Each cell corresponds to a stochastic differential equation defined by the gene regulatory network. Therefore, we need to track the gene expression dynamics of each individual cell. The number scheme is given below:

\begin{itemize}
	\item \textbf{input}
	~ The topology file of the gene regulatory network, containing gene nodes, regulatory relationships, and related parameters.
	\item \textbf{initialization} ~ Read the topology file and model parameters. Initialize the number of simulation cycles $\texttt{cycle} = 0$, set the cycle length $T = 50$, and set the initial time $t = 0$. Set the initial cell number $N$. For each cell $\Sigma = \{C_j(\vec{X}_j) : j=1,\cdots, N\}$,  set the initial condition for the gene expression level $\vec{X}_j(0)$.	
	\item \textbf{simulation process}
	\begin{itemize}
		\item[] \textbf{for} $\texttt{cycle} = 0$ to $\texttt{max}\_{\texttt{cycle}}$ \textbf{do}
		\begin{itemize}
		\item[] Copy the sytem $\Sigma'= \Sigma$.
		\item[] \textbf{for} all cells $C_j \in \Sigma$ \textbf{do}
			\begin{itemize}
			\item[-] \textbf{Gene regulation dynamics} ~ Solve the stochastic differential equation \eqref{eq:13} with the stochastic Runge-Kutta method for $0 \leq t \leq T$.
			\item[-] \textbf{Cell division}
			\\ ~ Generate two new cells $C_{j,1}(\vec{Y}_1)$ and $C_{j,2}(\vec{Y}_2)$, $\vec{Y}_l = (Y_{l,1}, \cdots, Y_{l,m})\ (l=1,2)$ store the initial gene expression state of the cell $l$.
			\\ ~ \textbf{for} $l=1$ to $2$ and $i=1$ to $m$ \textbf{do}
			\\ \mbox{}\quad Generate a parameter $\chi_i\sim \mathrm{Beta}(a, b)$;
			\\ \mbox{}\quad Reset the expression state $Y_{1,i}(0) = \chi_i X_{j,i}(T)$. 
			\\ \textbf{end~for}
			\\ Replace the cell $C_j$ in $\Sigma'$ with the new cells $C_{j,1}$ and $C_{j,2}$.
			\end{itemize}
			\item[] \textbf{end~for}
			\item[] $\mathtt{cycle} = \mathtt{cycle} +1$.
			\item[] Copy the system $\Sigma = \Sigma'$.
		\end{itemize}
		\item[] \textbf{end~for}
	\end{itemize}
\end{itemize}
	
\subsection*{B. Numerical scheme of the individual-cell-based simulation method}
	
\label{app:4}

The differential-integral equation \eqref{4'} can be solved numerically through the Euler method and numerical integration. However, it is expensive to calculate the numerical integration due to the high-dimensional epigenetic states. Hence, we usually do not solve the equation \eqref{4'} directly. We apply an individual-cell-based simulation method to simulate the cell cycling process.

In the model simulation, a multiple-cell system is represented as a collection of epigenetic states, and the individual-cell-based simulation tracks the behaviors of each cell according to its own epigenetic states. The sketch of the numerical scheme is given below.  
	
\begin{itemize}
\item \textbf{input}~ The parameters, kinetotype functions $\beta$, $\kappa$, $\mu$, and $\tau$, inheritance function $p(\vec{x}, \vec{y})$, and the time step $\Delta t$.
	\item \textbf{initialize}~ Set the time $t=0$, the initial cell number $N_0$, and all cells $\Sigma = \{[C_i(\vec{x})]_{i=1}^Q\}$. At the initial state, all cells are at the resting phase ($S = 0$), and the corresponding age at the proliferating phase is $a = 0$. Set the cell number at the resting phase as $N_q = N_0$ and the cell number at the proliferative phase as $N_p = 0$. 
	\item \textbf{simultion process}
	\begin{itemize}
		\item[] \textbf{for} $t$ from $0$ to $\texttt{T}_\texttt{end}$ with step $\Delta t$ \textbf{do}
		\item[] \quad\textbf{for} each cell in $\Sigma$ \textbf{do}
		\begin{itemize}
			\item[] \textbf{Kinetotype} ~Calculate the proliferation rate $\beta$, the apoptosis rate $\mu$, and the terminate differentiation rate $\kappa$.
			\item[] \textbf{Cell fate decision} ~Determine the cell fate during the time interval $(t, t+ \Delta t)$:
			\begin{itemize}
				\item[-] When the cell is at the resting phase, undergo terminal differentiation with a probability $\kappa \Delta t$, or enter the proliferation phase with a probability $\beta\Delta t$, or, if otherwise, remain unchanged. 
				\item[-] When a cell is at the proliferative phase, if the age $a < \tau$, the cell is either removed (through apoptosis) with a probability $\mu \Delta t$ or remains unchanged; if the age $a \geq \tau$, the cell undergoes mitosis and divides into two cells. 
			\end{itemize}
		\end{itemize}
			\item[] \quad\textbf{end for}

		\item[] \quad \textbf{system update}:  \textbf{for} each cell in $\Sigma$ \textbf{do}
		\begin{itemize}
		\item[-] If the cell fate is differentiation or apoptosis, remove the cell, and reduce the total cell number by 1, $N = N-1$. Accordingly, reduce the number of resting or proliferative cells: $N_q = N_q - 1$ or $N_p = N_p - 1$.
		\item[-]  If the cell fate is entering the proliferative state, set the cell state at proliferative ($S=1$) and the proliferating age $a=0$, and let $N_q = N_q - 1$, $N_p = N_p +1$, the total cell number $N$ is unchanged.
		\item[-]  If the cell is at the proliferative phase and remains unchanged, update the proliferative age $a = a + \Delta t$, the cell numbers $N_q$, $N_p$, and $N$ are unchanged. 
		\item[-] If the cell is under mitosis, it produces two daughter cells, and the epigenetic state of each daughter cell is determined according to the inheritance probability $p(\vec{x}, \vec{y})$. The proliferating age of each daughter cell is set as $a = 0$, and update the cell number: $N_q = N_q + 2$, $N_p = N_p - 1$, and $N= N+1$.
		\end{itemize} 
		\item[] \quad\textbf{end~for}
		\item[] \textbf{end~for}
	\end{itemize}
\end{itemize}
	
To determine the epigenetic state of daughter cells according to the inheritance probability $p(\vec{x}, \vec{y})$ defined by the combination gamma distribution \eqref{eq:29}, we applied the following numerical scheme:
\begin{itemize}
\item calculates the $\alpha_i, \psi_{i,j}, \eta_{i,j}$ according to the state of the mother cell, and the coefficients $a_{i,j}$ and $b_{i,j}$;
\item \textbf{for} $i=1$ to 2 \textbf{do}
\begin{itemize}
\item[-] generates a random number $q = \texttt{rand}()$;
\item[-] if $q < \alpha_i$, produce the epigenetic state of gene $i$ for daughter cell according to the first gamma distribution function $\mathrm{Gamma}(x_i; a_{i, 1}, b_{i,1})$, otherwise, produce the epigenetic state according to the second gamma distribution function $\mathrm{Gamma}(x_i; a_{i, 2}, b_{i,2})$.
\end{itemize} 
\item set the epigenetic state of the daughter cells.
\end{itemize}

In simulations, the cell number may increase to a very high number through cell division, which may cause a challenge issue in simulations. To overcome this issue, we applied a technique of \textit{downsampling}. The process of downsampling is often used in signaling analysis to reduce the data rate or the size of data. Here, we applied a similar technique to reduce the number of cells under simulation. 

We predefine a maximum number of cells to be simulated and stored ($N_{\mathrm{max}} = 10^6$ cells in the current simulation). Let $N_k = N_{k,q}  + N_{k,q}$ ($N_k \leq N_{\max}$) be the number of cells under simulation at step $k$, where $N_{k,q}$ represents the number of cells at the resting phase, and $N_{k, p}$ represent the number of cells at the proliferative phase. We first opened a temporary storage space for $N_{k, q} + 2 N_{k, p}$ cells (the maximum number of cells if all proliferative phase cells undergo mitosis). After performing cell fate decisions for each cell, we have potentially $N_{\mathrm{temp}} = N_{k,q}' + N_{k,p}'$ ($N_{\mathrm{temp}}\leq N_{k, q} + 2 N_{k, p}$) cells. If $N_{\mathrm{temp}} > N_{\mathrm{max}}$, we selected all cells ($N_{k,p}'$ cells) in the proliferative phase and at most $(N_{\mathrm{max}}-N_{k,q}')$ cells in the resting phase (select each cell with a probability $p = \max\{1,(N_{\mathrm{max}} - N_{k,q}')/N_{k,q}'\}$) to obtain $N_{\mathrm{next}}$ cells for the next step simulation; otherwise, we select all $N_{\mathrm{temp}}$ cells, and set $N_{\mathrm{next}} = N_{\mathrm{temp}}$. Finally, we stored the state of all selected $N_{\mathrm{next}}$ cells, let $f_{\mathrm{pro}, k} = N_{\mathrm{temp}}/N_k$ for the proliferation rate, and $N_{k+1} = N_{\mathrm{next}}$ for the number of cells simulated at step $k+1$. 

According to the approach of downsampling simulation, at step $k$, there are $N_k \leq N_{\mathrm{max}}$ cells under simulation, and the states of these cells are stored; the real cell number is given by
$$
N_{\mbox{real number at step}\ k} = N_0\prod_{i=0}^{k-1} f_{\mathrm{pro}, i}.
$$
This gives the cell number at step $k$.


\end{document}